\documentclass[conference]{IEEEtran}
\IEEEoverridecommandlockouts
\usepackage{cite}
\usepackage{amsmath,amssymb,amsfonts}
\usepackage{algorithmic}
\usepackage{graphicx}
\usepackage{textcomp}
\usepackage[nottoc]{tocbibind}
\usepackage{multirow}
\usepackage{xcolor}
\usepackage{booktabs}
\usepackage{threeparttable}
\usepackage{graphicx}
\usepackage{subfigure}
\def\BibTeX{{\rm B\kern-.05em{\sc i\kern-.025em b}\kern-.08em
    T\kern-.1667em\lower.7ex\hbox{E}\kern-.125emX}}
\begin{document}

\title{An Event-Driven Compressive Neuromorphic System for Cardiac Arrhythmia Detection \\
	\vspace{-0.5cm}
 \author{\IEEEauthorblockN{Jinbo Chen$^{1,2,*}$, Fengshi Tian$^{2,3,*}$, Jie Yang$^{2}$, and Mohamad Sawan$^{2}$, \IEEEmembership{Fellow, IEEE}}
 \IEEEauthorblockA{$^{1}$Zhejiang University, Hangzhou, Zhejiang, China 310058 \\
 $^{2}$CenBRAIN Lab., School of Engineering, Westlake University, Hangzhou, Zhejiang, China 310024 \\
 $^{3}$Department of ECE, HKUST, Hong Kong SAR, China \\
 \ Email: yangjie@westlake.edu.cn, sawan@westlake.edu.cn}
\vspace{-1.3cm}
\thanks{* Jinbo Chen and Fengshi Tian contributed equally to this work.}
}
}
\maketitle

\begin{abstract}

Wearable electrocardiograph (ECG) recording and processing systems have been developed to detect cardiac arrhythmia to help prevent heart attacks. Conventional wearable systems, however, suffer from high energy consumption at both circuit and system levels. To overcome the design challenges, this paper proposes an event-driven compressive ECG recording and neuromorphic processing system for cardiac arrhythmia detection. The proposed system achieves low power consumption and high arrhythmia detection accuracy via system level co-design with spike-based information representation. Event-driven level-crossing ADC (LC-ADC) is exploited in the recording system, which utilizes the sparsity of ECG signal to enable compressive recording and save ADC energy during the silent signal period. Meanwhile, the proposed spiking convolutional neural network (SCNN) based neuromorphic arrhythmia detection method is inherently compatible with the spike-based output of LC-ADC, hence realizing accurate detection and low energy consumption at system level. Simulation results show that the proposed system with 5-bit LC-ADC achieves 88.6\% reduction of sampled data points compared with Nyquist sampling in the MIT-BIH dataset, and 93.59\% arrhythmia detection accuracy with SCNN, demonstrating the compression ability of LC-ADC and the effectiveness of system level co-design with SCNN.

\end{abstract}


\vspace{-0.7cm}

\section{Introduction}

Cardiovascular disease (CVD) has been identified as one of the leading causes of human death based on the statistics from World Health Organization \cite{RN583}. Early detection of cardiac arrhythmia based on electrocardiograph (ECG) recording and processing could relieve the threat of CVD. Studies have shown that home monitoring systems with wearable ECG recording systems can effectively reduce heart failure hospitalization rates \cite{RN582,RN580,RN579,RN581}. Furthermore, recent research indicates a trend towards integrating ECG recording and processing into a wearable system for arrhythmia detection to achieve better energy efficiency and practicality at edge.\cite{RN35, RN315}. 

\begin{figure}[tb]
\centerline{\includegraphics[width=2.8in]{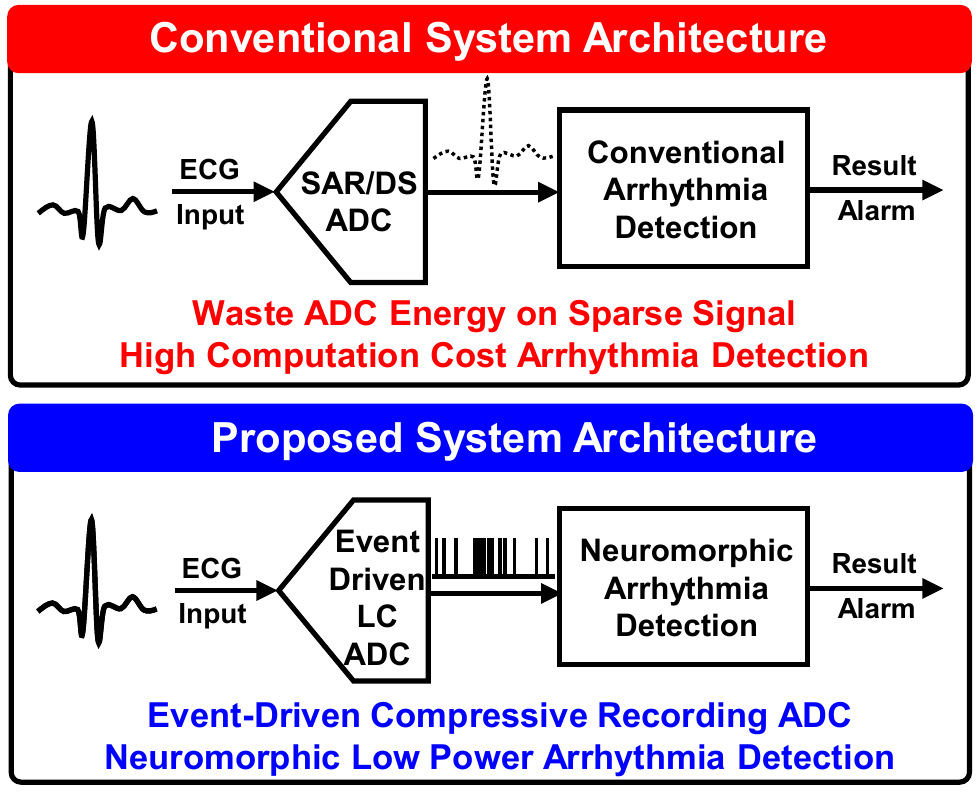}}
\vspace{-0.3cm}
\caption{Conventional system architecture (top); Proposed system architecture (bottom).}
\vspace{-0.7cm}
\label{fig_2system}
\end{figure}

The conventional architecture of wearable ECG recording and processing system is shown in Fig. \ref{fig_2system} (top). ECG signal after amplification and filtering is digitalized by SAR ADC or Delta-sigma ADC, and sent to arrhythmia detection engine to judge if there is an alarm or not.

Conventional system architecture faces critical design challenges. On the one hand, conventional recording architecture doesn’t take bio-signal features and sparsity into consideration, hence wasting ADC area and energy on unnecessary data \cite{RN236}. Moreover, after ADC digitalization, the information representation scheme is usually multi-bit discrete amplitude format \cite{RN35, RN315}, which is not computation-friendly for digital post-processing algorithms and leads to high computation power. Although some recent researches utilize specific ADC architecture that produces bit-streams based \cite{RN588} or pulse-based output \cite{RN587}, they are not co-designed with arrhythmia detection algorithms and thus have low system-level energy efficiency. 

On the other hand, conventional neural networks algorithms dedicated to arrhythmia detection and broad biomedical applications, such as artificial neural network (ANN), convolutional neural network (CNN) and binary neural network (BNN) \cite{RN590, RN16, RN591, RN587}, are not designed with the attention of the bio-signal recording architecture, thus dissipating computation power on unnecessary data. Moreover, these algorithms typically cannot process raw data from recording, and are hardware unfriendly. Although spike-based computing method has been proposed like \cite{RN589}, it is still not capable of directly processing raw data digitalized by ADC, which incurs extra hardware cost at system level.

To solve the aforementioned challenges, this paper proposes an event-driven compressive ECG recording and neuromorphic processing system for cardiac arrhythmia detection. The proposed system achieves low power and high detection accuracy by system level co-design and integrating event-driven compressive Level-Crossing ADC (LC-ADC) and Spiking Convolutional Neural Network (SCNN) based neuromorphic arrhythmia detection algorithm. The proposed system architecture is shown in Fig. \ref{fig_2system} (bottom). To the best of our knowledge, this is the first work that exploits LC-ADC and SCNN together to achieve spike-based information representation both in low power ECG recording and processing. The main contributions are listed as follows:

\vspace{-0.1cm}

\begin{itemize}
	\item Propose the LC-ADC based event-driven compressive ECG recording system to leverage the sparsity of ECG signal to achieve compressive recording and save ADC energy on silent signal.
	\item Propose the SCNN based neuromorphic processing method for arrhythmia detection, which is hardware-friendly and consumes low computation resources.
	\item The spike-based output of the proposed LC-ADC is inherently compatible with the proposed SCNN algorithm, thus realizing low energy consumption at system level.
	\item Simulation results show that the proposed system with 5-bit LC-ADC achieves 88.64\% data points reduction compared with Nyquist sampling in the MIT-BIH dataset, and 93.59\% arrhythmia detection accuracy with SCNN. This indicates the inherent compression ability of LC-ADC and the effectiveness of system level co-design with SCNN.
\vspace{-2pt}
\end{itemize}

The remaining sections of the paper are arranged as follows. Section II explains the architecture of the proposed system. Section III describes the implementation of the proposed system. Simulation results are presented in Section IV. Finally, Section V concludes the paper.

\section{Architecture of the Proposed System}

\subsection{Overall System Architecture}
\vspace{-0.1cm}

The proposed system investigates system level co-design to accomplish low power event-driven compressive ECG recording and neuromorphic processing for high arrhythmia detection accuracy by the cooperation of LC-ADC and SCNN algorithm. Fig. \ref{fig_2system} (bottom) presents the proposed system architecture. The ECG signal is converted into digital spikes by LC-ADC that utilizes level-crossing sampling with signal compression ability. Therefore, the number of sampled data points will be significantly reduced compared with Nyquist sampling. Moreover, the digital spikes are inherently compatible with the proposed SCNN based neuromorphic arrhythmia detection algorithm, which eliminates the hardware cost of extra spike encoders in conventional SCNN implementations. This enables the presented SCNN to directly process the spike-based output of LC-ADC with low computation complexity and power consumption.

\vspace{-0.2cm}

\subsection{Level-Crossing ADC Based Event-Driven Compressive ECG Recording}

\vspace{-0.1cm}

\begin{figure}[tb]

	\centerline{\includegraphics[width=3.5in]{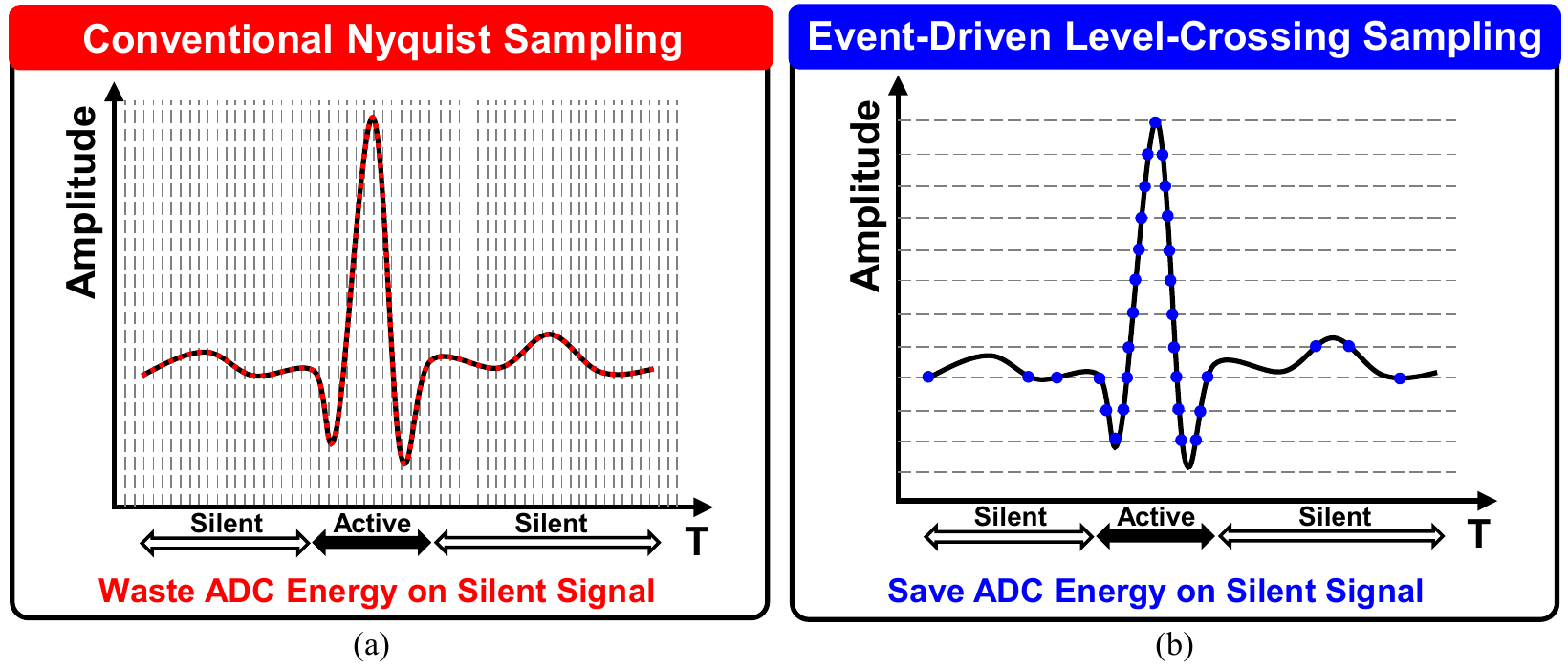}}
	\vspace{-0.3cm}
	\caption{(a) Nyquist sampling; (b) Level-crossing sampling.}
	\vspace{-0.4cm}
	\label{fig_lcsamping}
\end{figure}

Features used for arrhythmia detection are sparse within ECG signals. As displayed in Fig. \ref{fig_lcsamping}, most ECG signals are silent and irrelevant to arrhythmia detection. However, conventional uniform Nyquist sampling as illustrated in Fig. \ref{fig_lcsamping} (a) constantly samples ECG signal without considering its sparsity, thus wasting ADC energy during silent period. By contrast, LC-ADC is a nonuniform sampling frequency data converter. It could utilize the sparsity of ECG, and only sample signal when predefined levels are crossed. Fig. \ref{fig_lcsamping} presents the difference between conventional Nyquist sampling and event-driven level-crossing sampling. As shown in Fig. \ref{fig_lcsamping} (b), there are no sampling points when the signal is silent and does not cross levels, which reduces the total number of sampled data points and thus saves ADC energy.

\vspace{-0.3cm}

\begin{figure}[tb]
	\centerline{\includegraphics[width=2.3in]{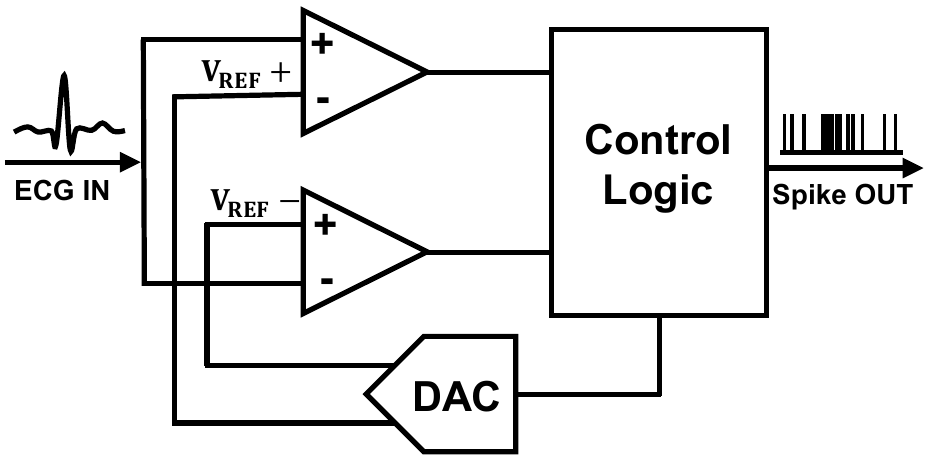}}
	\vspace{-0.3cm}
	\caption{Basic LC-ADC architecture.}
	\vspace{-0.6cm}
	\label{fig_lcadc}
\end{figure}

\vspace{+0.3cm}
Fig. \ref{fig_lcadc} illustrates the basic LC-ADC architecture. The ECG voltage range is divided into predefined quantization levels that are represented by $V_{ref}+$ and $V_{ref}-$. The numerical distance of two consecutive quantization levels is called the LSB, which is calculated from the equation below: 
\vspace{-0.1cm}
\begin{equation}
LSB=\frac{A_{FS}}{2^{M}}
\label{equ:LCADC}
\end{equation}
where $A_{FS}$ refers to the voltage amplitude range of the input signal. The resolution bit of LC-ADC is M, producing ${2^{M}}$ quantization levels. The LC-ADC samples input signal and produces spike output when the quantization levels are up or down crossed. 
\vspace{-0.2cm}

\subsection{SCNN Based Energy Efficient Neuromorphic Arrhythmia Detection}
\vspace{-0.1cm}

\begin{figure}[t]
	\begin{center}
		\includegraphics[width=3.5in]{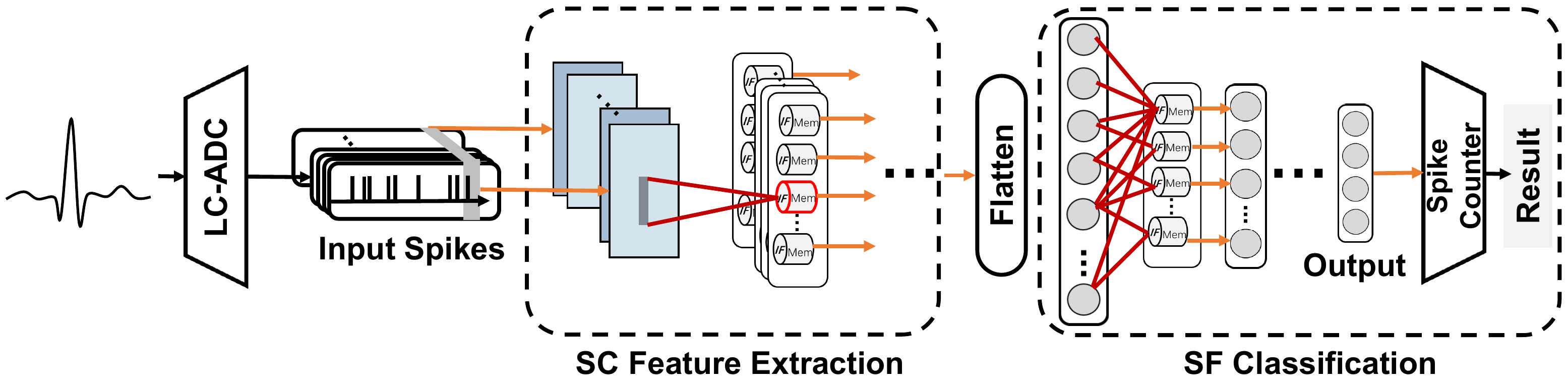}
		\vspace{-0.7cm}
		\caption{Structure of SCNN.}
		\vspace{-0.8cm}
		\label{fig:SCNN}
	\end{center}
\end{figure}

As shown in Fig. \ref{fig:SCNN}, the computing strategy of the proposed SCNN involves two phases. The spiking convolution (SC) feature extraction phase consists of 1-D convolution and max-pooling kernels for feature extractions, where the 1-D convolution kernels extract key features from data and the max-pooling kernels maintain the most useful information. The SC phase receives spike data from LC-ADC directly and sends single-bit dataflow to the next phase. The single-bit dataflow from the SC feature extraction phase is then flattened into a single-dimension vector. The spiking fully-connection (SF) classification phase receives the single-dimension vector from the SC phase and adopts fully-connected layers to operate classifications and tell the result of arrhythmia detection. 

Different from CNN which uses rectified linear unit (ReLU) activation neurons, SCNN adopts leaky integrate and fire (LIF) neuromorphic neuron model to imitate the functions of real neurons \cite{RN593}. The LIF neurons can receive multiple spike inputs on the time domain and fire multiple output spikes in one processing procedure. $V_{mem}(t)$ and $V_{reset}$ stand for the membrane potentials and the reset potentials of such neurons at the time step t, respectively. When there exist input spikes, $V_{mem}(t)$ increases, but later returns to $V_{reset}$ after firing a spike at the output. When there exists no input spike, the membrane potential $V_{mem}(t)$ decreases by a leaky voltage $\bigtriangleup V$. The updating process at every time step is described in Eq. \eqref{equ:SCNNupdate}

\vspace{-0.6cm}
\begin{equation}
V_{mem,i}(t)=V_{mem,i}(t-1)+\sum w i j \sum \delta j(t-t j)-\Delta V
\label{equ:SCNNupdate}
\end{equation}
\vspace{-0.5cm}

At every time step, the time-dependent spike data flows through the whole SCNN and some or all or none of the output neurons are activated to fire spikes. The spike counter in Fig. \ref{fig:SCNN} records the firing times of every output neuron across all the time steps. Then the greatest one determines the classification result. The proposed SCNN does not require multiplication operations as all data flowing between layers is one-bit spike data (0 or 1). The required memory for data storage is also hugely decreased. Therefore, much computation and memory cost are reduced, making the proposed arrhythmia detection method energy efficient.

\section{Implementation of the Proposed System}

\subsection{ECG Dataset}
\vspace{-0.1cm}

The MIT-BIH arrhythmia dataset is used to evaluate the performance on ECG based arrhythmia detection tasks. It includes 48 half-hour excerpts from two-channel ambulatory ECG recordings. These recordings were collected between 1975 and 1979 from 47 subjects analyzed by the MIT BIH Arrhythmia Laboratory. Among the total 48 recordings, 23 are selected at random from a set of 4 thousand 24-hour ambulatory ECG recordings that are obtained from a population consisting of 60\% inpatients  and 40\% outpatients at Boston's Beth Israel Hospital. From the same set, the additional 25 recordings were chosen to involve less ordinary but still clinically important arrhythmias that couldn't be adequately reflected in a randomly small sample \cite{RN594}.

The recordings were digitalized at a sampling rate of 360Hz with 11-bit resolution and a range of 10 mV. ECG recordings can be classified into five categories according to the Association for the Advancement of Medical Instrumentation (AAMI) standard: non ectopic beat (N), supraventricular ectopic beat (SVEB), ventricular ectopic beat (VEB), fusion beat (F) and unknown beat (Q). In the MIT-BIH dataset, there are 90081 N beats, 2781 SVEB beats, 7008 VEB beats, 802 F beats and 15 Q beats in total. In this work, except for Q beats, 800 beats are selected randomly from N, SVEB, VEB and F, respectively to overcome the problem of data unbalance. Therefore, there are 3200 samples used in total, which are separated into training set and testing set randomly with the ratio of 4:1.
\vspace{-0.2cm}

\subsection{Configurable Modelling of LC-ADC}
\vspace{-0.1cm}

The LC-ADC is modeled and simulated using MATLAB. The input signal is from MIT-BIH arrhythmia dataset, which is digitalized by a Nyquist ADC with aforementioned specifications\cite{RN586}. To convert the dataset into level-crossing sampling spikes, an LC-ADC in \cite{RN585} is exploited with the clock frequency of 360 Hz. The $A_{FS}$ in Eq. \eqref{equ:LCADC} is set as 10mV considering the voltage range of MIT-BIH dataset. The resolution bit M is varied from 5-bit to 7-bit. During the sampling process, the first Nyquist sampled point of every record is chosen as a reference point, and the value of subsequent Nyquist sampled points is consecutively compared with the reference point. The LC-ADC will produce output spikes once the variation between the current point and the reference point is equivalent to or larger than one LSB. Then, the previous reference point is changed to the present point. This procedure is iterated until  all points' comparison is completed. 

The simulated LC-ADC produces two types of output spikes, i.e., REQ and DIR. The REQ spikes are generated when the predefined levels are crossed. While the DIR spikes are generated only when up crossing of input signal happens, which stores the direction information of input signal. Considering the data format requirements of SCNN, the REQ and DIR spikes of each ECG sample are merged into one row data comprising positive and negative spikes to denote both amplitude and direction information.
\vspace{-0.1cm}

\subsection{Adaptive SCNN and LIF Neuron}
\vspace{-0.1cm}

An adaptive network structure for SCNN is proposed in this work. The adaptive neural network topology adopts multi-layer channel-wise single-dimension convolution and max-pooling kernels for feature extractions, which can be adapted to obtain the best performance. The proposed SCNN comprises 5 spiking convolution layers along with 2 max-pooling layers in the SC feature extraction phase and 2 spiking fully-connected layers plus a spike counter layer in the SF classification phase. A LIF neuron layer is also connected after each spiking convolution layer and each spiking full-connected layer. The proposed SCNN is trained via backpropagation \cite{RN595}.

For LIF neurons, the reset potential, threshold voltage and leaky voltage are the adaptive parameters. In this work, all the LIF neurons share the same parameter settings for the convenience of hardware implementation. Inspired by natural behavior patterns of biological neurons, a self-modulated function is employed on the adaptive LIF neuron model in this work. Such function can prevent the LIF neurons from being constantly active or inactive to avoid overfitting of the SCNN, which may cause huge degradation in performance. 

\section{Simulation Results}

\subsection{Compression Performance of LC-ADC}
\vspace{-0.1cm}

\begin{figure}[tb]
	\centerline{\includegraphics[width=3.1in]{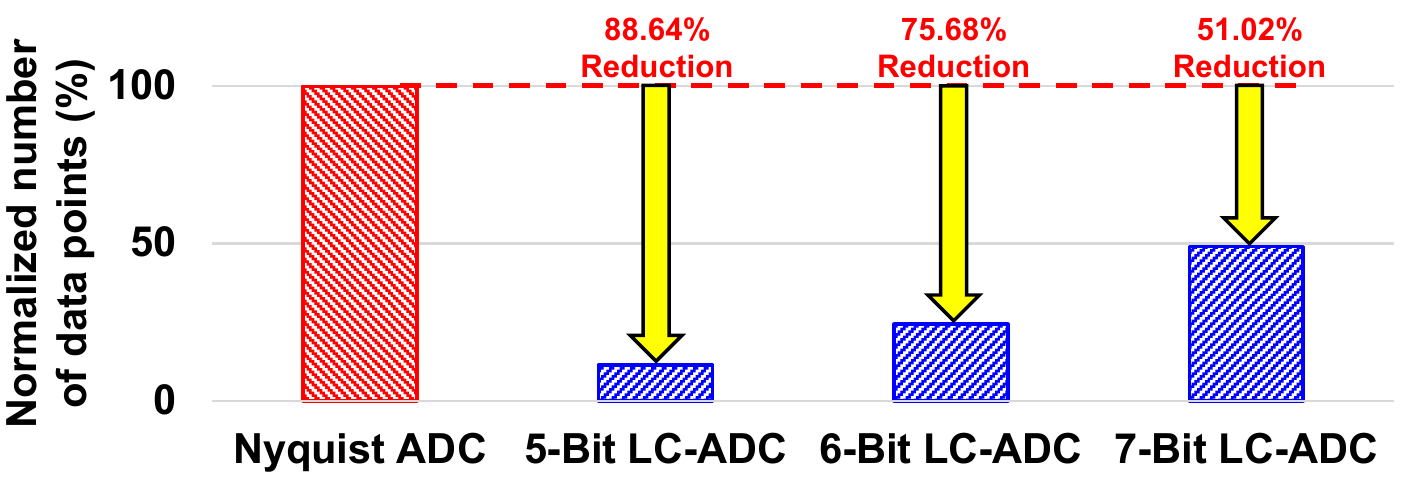}}
	\vspace{-0.3cm}
	\caption{Compression performance of LC-ADC.}
	\vspace{-0cm}
	\label{fig_compression}
\end{figure}

Fig. \ref{fig_compression} shows the compression performance of LC-ADC from 5-bit to 7-bit. Compared with Nyquist sampling in MIT-BIH dataset, the reduction of normalized number of data points is 88.64\% for 5-bit LC-ADC, 75.68\% for 6-bit LC-ADC, and 51.02\% for 7-bit LC-ADC. The reduction of data points demonstrates the inherent compression ability of LC-ADC. This is achieved by exploiting the sparsity of ECG signal and not digitalizing the signal during its silent period, which will save ADC energy from hardware perspective \cite{RN230}. Fig. \ref{fig_compression} also explains that the compression ability is negative correlated with the value of LC-ADC resolution bit.
\vspace{-0.2cm}

\subsection{Detection Performance of SCNN}
\vspace{-0.1cm}

\begin{figure}[tb]
	\centerline{\includegraphics[width=3.1in]{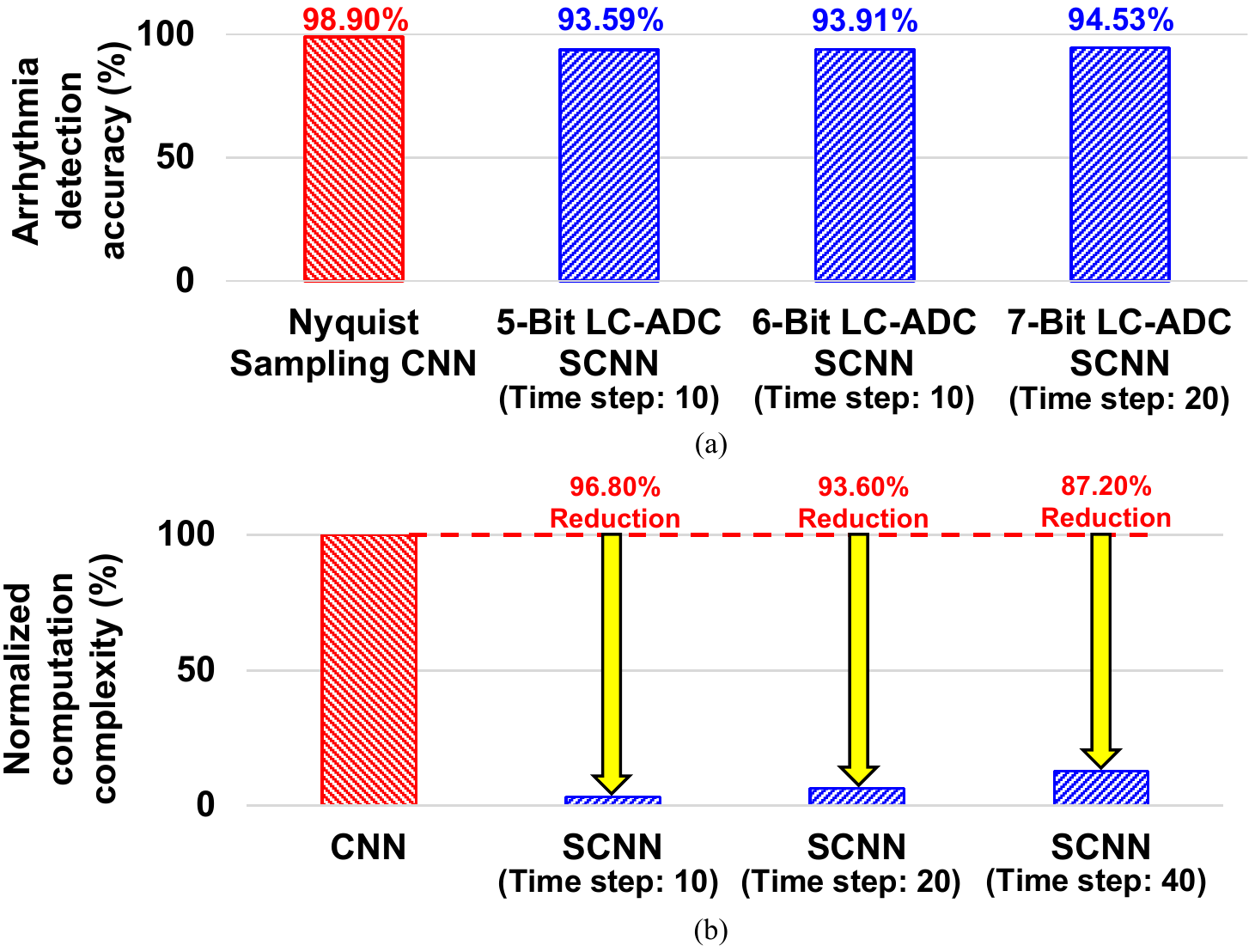}}
	\vspace{-0.3cm}
	\caption{Performance comparison of CNN and SCNN. (a) Arrhythmia detection accuracy; (b) Normalized computation complexity.}
	\vspace{-0.6cm}
	\label{fig_detection_computation}
\end{figure}

Fig. \ref{fig_detection_computation} (a) shows the arrhythmia detection performance of the proposed SCNN with various n-bit LC-ADC sampling output and comparison with CNN of the same structure. The input data of CNN is the Nyquist sampling value while the SCNN input data is spike sequence from LC-ADC. The SCNN keeps an accuracy of over 93.5\%. 

To demonstrate the computation reduction of the proposed SCNN compared with CNN-based methods, the computation complexity metric is calculated. The computation complexity of CNN is presented in Eq. \eqref{equ:TCCNN} and the computation complexity of SCNN is explained in Eq. \eqref{equ:TCSCNN}.

\vspace{-0.5cm}
\begin{equation}
\begin{array}{r}
TC_{CNN}=\sum M_{H} M_{W}\left(K_{H} K_{W}+K_{H}+K_{W}-1\right) \\ \times C_{in} C_{out} \times Ops \times bit +\sum N_{in} N_{out} \times Ops \times bit 
\end{array}
$$
\label{equ:TCCNN}
\vspace{-0.7cm}
$$\end{equation}
\vspace{-1.2cm}

\begin{equation}
\begin{array}{c}
TC_{SCNN}=\sum M_{H} M_{W}\left(K_{H}+K_{W}-1\right) C_{in} C_{out} \\ \times Ops \times bit \times t +\sum N_{in} N_{out} \times Ops \times bit \times t
\end{array}
$$
\label{equ:TCSCNN}
\vspace{-0.8cm}
$$\end{equation}
where TC, M, N, K, C, H, and W stand for Time Complexity, Feature Map Size, Neuron Count in FC Layers, Kernel, Channel, Height and Width; Ops is needed cycles in one computing operation (1 for adding, 1 for conditional branching and 10 for multiplication); bit stands for number of bits of data flows between two layers in NN; t is number of time steps. For the proposed SCNN, the only needed operation is adding and conditional branching for judging if the spike output should be emitted or not, while CNN needs MAC operations. SCNN uses one bit 0 or 1 as data flow while CNN uses 32-bit floating points. As shown in Fig. \ref{fig_detection_computation} (b), the proposed SCNN manages to reduce 96.8\% computation complexity with less than 5\% loss in accuracy compared to CNN.
\vspace{-0.2cm}

\subsection{Sensitivity Analysis of the Proposed System}
\vspace{-0.1cm}

\begin{figure}[tb]
	\centerline{\includegraphics[width=3.5in]{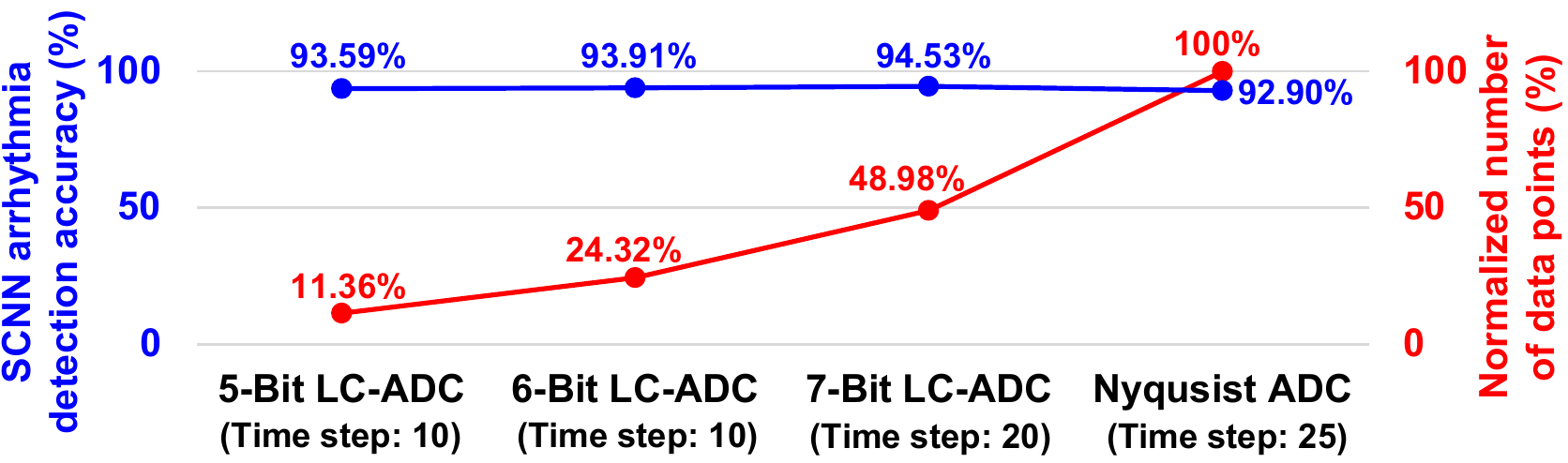}}
	\vspace{-0.3cm}
	\caption{Sensitivity analysis of the proposed system.}
	\vspace{-0.5cm}
	\label{fig_sensitivity_analysis}
\end{figure}

Fig. \ref{fig_sensitivity_analysis} presents the compression and detection performance with different resolution bits of LC-ADC and various time steps. As illustrated in Fig. \ref{fig_sensitivity_analysis}, the proposed SCNN algorithm achieves similar detection performance with different resolution bits of LC-ADC. 5-bit LC-ADC is the optimal choice because it offers the highest data compression ability. Moreover, it can be seen from Fig. \ref{fig_sensitivity_analysis} that the average detection accuracy of SCNN with LC-ADC is 94.01\%, which is higher than 92.9\% of SCNN with Nyquist ADC in the MIT-BIH dataset. This further demonstrates the effectiveness of the proposed system level co-design with spike-based information representation. Table I compares this work with other works.
\vspace{-0.1cm}

\begin{table}[t]
	\label{table_1}
	\caption{Performance comparison of state-of-art works.}
	\vspace{-0.3cm}
	\resizebox{\linewidth}{!}{%
\begin{tabular}{|c|c|c|c|c|c|}
	\hline
	\textbf{Specifications} &
	\textbf{\begin{tabular}[c]{@{}c@{}}BioCAS\\ 2019 \cite{RN597}\end{tabular}} &
	\textbf{\begin{tabular}[c]{@{}c@{}}ISCAS\\ 2021 \cite{RN584}\end{tabular}} &
	\textbf{\begin{tabular}[c]{@{}c@{}}DATE\\ 2018 \cite{RN599}\end{tabular}} &
	\textbf{\begin{tabular}[c]{@{}c@{}}JBHI\\ 2014 \cite{RN598}\end{tabular}} &
	\textbf{This Work} \\ \hline
	\begin{tabular}[c]{@{}c@{}}ADC\\ Architecture\end{tabular} &
	\begin{tabular}[c]{@{}c@{}}11-Bit\\ SAR ADC\end{tabular} &
	\begin{tabular}[c]{@{}c@{}}12-Bit\\ SAR ADC\end{tabular} &
	\begin{tabular}[c]{@{}c@{}}11-Bit\\ SAR ADC\end{tabular} &
	\begin{tabular}[c]{@{}c@{}}12-Bit\\ Sigma Delta\\ Modulator\end{tabular} &
	\begin{tabular}[c]{@{}c@{}}5-Bit\\ Level-Crossing\\ ADC\end{tabular} \\ \hline
	\begin{tabular}[c]{@{}c@{}}ADC Data\\ Compression\end{tabular}               & No      & No      & No      & No      & Yes \\ \hline
	\begin{tabular}[c]{@{}c@{}}ADC Output\\ Information\\ Representation\end{tabular} &
	\begin{tabular}[c]{@{}c@{}}Multi-Bit\\ Discrete\\ Amplitude\end{tabular} &
	\begin{tabular}[c]{@{}c@{}}Multi-Bit\\ Discrete\\ Amplitude\end{tabular} &
	\begin{tabular}[c]{@{}c@{}}Multi-Bit\\ Discrete\\ Amplitude\end{tabular} &
	\begin{tabular}[c]{@{}c@{}}Multi-Bit\\ Discrete\\ Amplitude\end{tabular} &
	\begin{tabular}[c]{@{}c@{}}Spike\\ Sequence\end{tabular} \\ \hline
	\begin{tabular}[c]{@{}c@{}}Detection\\ Algorithm\end{tabular} &
	\begin{tabular}[c]{@{}c@{}}Multi-Level\\ SVM\end{tabular} &
	\begin{tabular}[c]{@{}c@{}}Ternary\\ Neural\\ Network\end{tabular} &
	\begin{tabular}[c]{@{}c@{}}Threshold-\\ Based\end{tabular} &
	\begin{tabular}[c]{@{}c@{}}Wavelet\\ Transform\end{tabular} &
	SCNN \\ \hline
	\begin{tabular}[c]{@{}c@{}}Algorithm\\ Compatibility\\ With ADC\end{tabular} & No      & No      & No      & No      & Yes     \\ \hline
\end{tabular}
	}
	\vspace{-0.6cm}
	\end{table}
	
\vspace{-0.1cm}
\section{Conclusion}

This paper proposes an event-driven compressive ECG recording and neuromorphic processing system for cardiac arrhythmia detection that reduces the number of sampled data points by 88.64\% compared with Nyquist sampling in the MIT-BIH dataset and accomplishes 93.59\% arrhythmia detection accuracy. The presented system provides low power and high detection accuracy by designing the system architecture where spike-based information flows from recording to processing modules. The reduction of sampled data points results from an event-driven 5-bit LC-ADC that utilizes the sparsity of the ECG signal to achieve compressive recording and saves ADC energy during the silent ECG signal period. Additionally, the spike output of the LC-ADC can be inherently applied to the proposed SCNN, resulting in a detection accuracy comparable to that of CNN while requiring significantly less computation complexity. Overall, the simulation results demonstrate the LC-ADC's compression ability and system-level power reduction by co-design with SCNN.

\vspace{-0.2cm}

\section*{Acknowledgments}
This work is funded by start-up funds from Westlake University, Zhejiang Key R \& D Program project No. 2021C03002, and Zhejiang Leading Innovative and Entrepreneur Team Introduction Program No. 2020R01005.

\clearpage

{\small{
\bibliographystyle{IEEEtran}
\bibliography{reference.bib}}
}
\end{document}